
\magnification=1000
\baselineskip=13pt
\font\tipo=cmr10
\font\it=cmti10 scaled 1000
\font\tit=cmbx10 

\font\ref=cmti9
\hoffset=1.5truecm
\voffset=1.1truecm
\hsize=5.5truein 
\vsize=7.7truein

\tipo

\

\vskip-3truecm

\rightline{ICN-UNAM, Mexico, November 16, 1993.}

\

\vskip2truecm

\centerline{\tit $B \wedge F$ THEORY AND FLAT
SPACETIMES\footnote{$^\ast$}{\tipo This research is supported in part
by the National Science Foundation, Grant No. PHY 89-04035, by
CONACyT Grant No. 400349-1714E (Mexico), and by the Association
G\'en\'erale pour la Coop\'eration et le D\'eveloppement
(Belgium).}}

\vskip1cm

\baselineskip12pt
\centerline{\tipo Henri Waelbroeck}

\centerline{Instituto de Ciencias Nucleares,}

\centerline{Circuito Exterior, C.U.}

\centerline{A. Postal 70-543}

\centerline{M\'exico 04510, D.F., M\'exico}

\

\tipo
\baselineskip10pt
{\leftskip=1.5pc\rightskip=1.5pc\smallskip\noindent
Abstract: We propose a reduced constrained Hamiltonian formalism
for the exactly soluble $B \wedge F$ theory of flat connections
and closed two-forms over manifolds with topology $\Sigma^3 \times
(0,1)$. The reduced phase space variables are the holonomies of
a flat connection for loops which form a basis of the first
homotopy group $\pi_1(\Sigma^3)$, and elements of the second
cohomology group of $\Sigma^3$ with value in the Lie algebra
$L(G)$. When $G=SO(3,1)$, and if the two-form can be expressed as
$B= e\wedge e$, for some vierbein field $e$, then the variables
represent a flat spacetime. This is not always possible: We show
that the solutions of the theory generally represent
spacetimes with ``global torsion''. We describe the dynamical
evolution of spacetimes with and without global torsion, and
classify the flat spacetimes which admit a locally homogeneous
foliation, following Thurston's classification of geometric
structures.\smallskip}

\tipo

\

\tipo
\baselineskip13pt
\noindent
1. INTRODUCTION

\

The $B \wedge F$ theory was first considered by Horowitz [1] as
an example of an exactly soluble theory in four dimensions,
analogous to the Chern-Simons formulation of (2+1)-dimensional
gravity [2][3]. The set of solutions was shown to be related to
equivalence classes of flat $SO(3,1)$ connections and closed
two-forms. When the four-manifold has the topology $\Sigma^3 \times
(0,1)$, where $\Sigma^3$ is compact and orientable, there is a
natural symplectic structure which is related to the Poincar\'e
duality between the first and second homology groups of
$\Sigma^3$: Roughly speaking, the flat connections are labeled by
their holonomies around loops of $Z_1(\Sigma^3)$, and the
two-forms are labeled by their integrals over elements of
$Z_2(\Sigma^3)$. The symplectic structure on the set of
holonomies and integrated two-forms would then be derived from
the Poincar\'e duality between $H_1 (\Sigma^3) = Z_1 (\Sigma^3)/
B_1 (\Sigma^3)$ and $H_2 (\Sigma^3) = Z_2
(\Sigma^3)/B_2(\Sigma^3)$.

The purpose of this article is to elucidate further the physical
content of this theory. We will first postulate a reduced
constrained Hamiltonian formalism [4] which exploits the
Poincar\'e duality explicitly, and can be solved for the
dynamical evolution of the reduced phase space variables. Next,
we will examine the relation between the solutions of two-form
gravity and flat spacetimes, including a classification of flat
spacetimes which admit a locally homogeneous foliation.

The article is organized as follows. We will begin with a review
of the $B\wedge F$ field theory, derive the constrained
Hamiltonian system for the reduced variables, and examine the
dynamical evolution (Section 2). In Section 3, we will show that
these solutions generally represent spacetimes  with ``global
torsion'', and propose vanishing torsion constraints. Finally, we
will construct all flat spacetimes which admit a locally
homogeneous slicing, by constructing the representations $\rho:
\pi_1 (\Sigma^3) \to ISO(3,1)$ for each of Thurston's eight
geometric structures.

\

\tipo
\baselineskip13pt
\noindent
2. TWO-FORMS AND FLAT CONNECTIONS

\

\noindent
2.1 The $B\wedge F$ Field Theory

\

The $B \wedge F$ theory describes a gauge connection $A$ in $G$,
and a two-form $B$ with values in the Lie algebra $L(G)$, with the
action

$$S = 6 \int_M tr (B\wedge F) \eqno(2.1)$$

The field equations which derive from variations of $B$ and $A$
are, respectively, $F = 0$ and $D\wedge B = 0$, where $F$ is the
field strength and $D = \partial + A$ is the gauge-covariant
derivative. This action is invariant under the gauge
transformations $\delta A = D\tau, \delta B = [B, \tau]$, where $\tau$
is an arbitrary field on $M$ with values in the Lie algebra
$L(G)$, and under the translation of $B$ by an exact form,
$\delta B = D\wedge v$. When $A$ satisfies its field equation,
$D^2 = F = 0$, there is a cohomology of covariantly closed
two-forms, modulo translations by covariantly exact forms. This
implies that the physical information carried by $B$ can be
represented by a map form the second homology group
$H_2(\Sigma^3)$ to the Lie algebra $L(G)$: When $G$ is abelian,
this map is simply the surface integral of the two-form over any
representative of $H_2(\Sigma^3)$ in $Z_2(\Sigma^3)$.

If one sets $G=SO(3,1)$ and $B^{ab} = e^a \wedge e^b$, the action
(2.1) becomes the Palatini action for vacuum gravity,

$$S_P = \int (e^a \wedge e^b \wedge F^{cd})
\epsilon_{abcd}\eqno(2.2)$$

\noindent
The relation between the $B\wedge F$ theory and gravity will be
examined further in Sect. 3.

As Horowitz has pointed out, if $M = \Sigma^3 \times (0,1)$ the
reduced phase space has a canonical structure which is related to
the Poincar\'e duality between $H_1(\Sigma^3)$ and
$H_2(\Sigma^3)$: The degrees of freedom of the connection are
associated to non-contractible loops, while the inequivalent
closed two-forms differ by the values of their integrals over
non-contractible surfaces of $Z_2(\Sigma^3)$. Poincar\'e duality
suggests that if the former are configuration space variables,
then the latter should be the corresponding momentum variables.
Before we can make this idea more explicit (Sect. 2.2), we first
need to put the action (2.1) in canonical form.

Assuming that $M = \Sigma^3 \times (0,1)$, one can separate the
coordinates into ``spatial'' (i,j,k) and ``time'' (o) components,
and write the action in the form

$$S = \int dt \int d\Sigma (\dot A^A_{[i} B_{jk]A} - F^A_{[ij}
B_{k]o A} + A^A_o D_{[i} B_{jk]A})\eqno(2.3)$$

\noindent
(We will assume that $G$ admits a non-degenerate Cartan metric;
the index $A$ is in the adjoint representation).

The canonical formalism is obtained by computing $\pi = {\partial
L \over {\partial \dot A}}$ and $\Pi = {\partial L \over
{\partial \dot B}}$, and performing the Legendre transformation.
One finds

$$[A_i{}^A , B_{jk B}] = \delta^A_B \Sigma_{ijk}\eqno(2.4)$$

$$\pi^0_A = {\partial L\over{\partial \dot A_0{}^A}} \approx
0\eqno(2.5)$$

$$\Pi^{ko}{}_A = {\partial L\over {\partial \dot B^A{}_{ko}}}
\approx 0 \eqno(2.6)$$

$$H = 3 \int_\Sigma \left( F^A_{[ij} B_{k]oA} + A_o{}^A D_{[i}
B_{jk]A} \right) d\Sigma \eqno(2.7)$$

\noindent
and the secondary constraints [4], which express the consistency
of (2.5) and (2.6) with time evolution, are

$$\dot \Pi^0_A \approx 0 \Rightarrow D_{[i} B_{jk]A} \approx 0
\eqno(2.8)$$

$$\dot\Pi^{ko}_A \approx 0 \Rightarrow F^A_{ij} \approx 0
\eqno(2.9)$$

These constraints state that $B^A_{jk}$ is a covariantly closed
form on $\Sigma^3$, and that $A^A_i$ (the connection resctricted
to $\Sigma^3$) is flat. Note that, given (2.8) and (2.9),

$$H \approx 0. \eqno(2.10)$$

\noindent
The flow generated by $H$ is

$$[H, A_i{}^A] = D_i A_0{}^A \eqno(2.11)$$

$$[H, B_{ij A}] = D_{[i} B_{k]0 A} + [B_{ij} , A_0]_A
\eqno(2.12)$$

\noindent
As Horowitz noted, diffeomorphisms are part of the symmetry
group when the constraints are satisfied. For any vector field
$\xi$,

$$L_\xi A_j^A = D_{[j} \left(A_{i]}^A \xi^i \right) + \xi^i
F_{ij}{}{}^A, \eqno(2.13)$$

\noindent
which reduces to the form (2.11) when $F_{ij}{}{}^A = 0.$

\

\noindent
2.2 The reduced theory

\

Since the fundamental group is presented by a finite basis
$\{\gamma(\mu), \mu =$ \break $1, \cdots , dim H_1(\Sigma^3)\}$, the
representations $\rho : \pi_1 (\Sigma^3) \to SO(3,1)$ are
parametri-\break zed by the holonomies for a set of basis loops:

$$M(\mu) = P exp \biggl\{-i \oint_{\gamma(\mu)} A_i dl^i\biggr\} ,
\eqno(2.14)$$

\noindent
where $P$ denotes the usual path ordering of the exponential. The other
half of the phase space is parametrized by elements of the second
cohomology group with values in the Lie algebra $L(G)$. In de
Rahm cohomology, the equivalence classes of closed two-forms modulo
translations by exact forms can be parametrized by their
integrals over basis surfaces $\sigma(\mu)$, where each
$\sigma(\mu)$ intersects only the basis loop $\gamma(\mu)$ only
at the base-point of $\pi_1(\Sigma^3)$. Generalizing this idea to
the covariant cohomology defined by $D = \partial + A$ when $F =
0$, one would naively consider the integrals

$$B^A(\mu) = \oint_{\sigma(\mu)} P^A{}_B (x) B_{ij}{}{}^B
(x) dx^i \wedge dx^j , \eqno(2.15)$$

\noindent
where $P^A{}_B(x)$ parallel-transports the index $B$ from
the fiber at $x$ to the fiber at the base-point $P$ of the
fundamental group $\pi_1(\Sigma^3)$.

The key problem is that this parallel-transport is not
well-defined, since different paths from $x$ to $P$ can
differ by non-trivial elements of $\pi_1(\Sigma^3)$. To define
uniquely the parallel-transport between points of $\Sigma^3$, one
can instead remove the $dim H_2(\Sigma^3)$ basis surfaces
$\sigma(\mu) \epsilon Z_2(\Sigma^3)$ to form a topologically
trivial open manifold $\Sigma^3_0$. The integrated two-form is then
defined by approaching the removed surface from one side or
the other:

$$B^A(\mu) = \lim_{\epsilon\to 0} \oint_{\sigma_{+\epsilon(\mu)}}
P^A{}_B(x) B_{ij}^B (x) dx^i \wedge
dx^j \eqno(2.16)$$

$$B^A(-\mu) = \lim_{\epsilon\to 0} \oint_{\sigma_{-\epsilon(\mu)}}
P^A{}_B(x) B^B_{ij}(x) dx^i \wedge
dx^j\eqno(2.17)$$

The difference between these expressions is precisely the
parallel-trans-\break port around the loop $\gamma(\mu)$, and we have

$$B^A(\mu) = M^A{}_B (\mu) B^B(-\mu)\eqno(2.18)$$

\noindent
To resolve the ambiguity in parallel-transport, we have been
forced to introduce a dependence on the surfaces $\sigma(\mu)$.
One can show that it is impossible to achieve {\it both}
gauge-invariance and surface-invariance within a homology class,
for any choice of parallel-trans-port $P^A{}_B (x)$. This
fact appears to be related to Teitelboim's proof that it is
impossible to achieve a reparametrization-invariant ordering of
the expression

$$P exp \left( \oint B_{ij} dx^i \wedge dx^j \right),
\eqno(2.19)$$

\noindent
analogous to the holonomy (2.14)[5]. Note
that the logarithm of Teitelboim's two-form ``holonomy'' can be
written in the form (2.15), but taking $P^A{}_B (x)$ as a
functional of $B$, rather than $A$.

The expression (2.16) is coordinate-invariant if the surfaces
$\sigma(\mu)$ are chosen in a coordi-nate-independent way. For
example, one might choose a maximal surface among the surfaces
$\sigma(\mu) \epsilon Z_2(\Sigma^3)$ in a given homology class
and which contain $P$. Any intrinsic criterion which selects a
unique representative in $Z_2(\Sigma^3)$ of each element of
$H_2(\Sigma^3)$, leads to an acceptable definition of the global
variables $B^A(\mu)$. For example, in Sect. 3, we will choose a
piecewise geodesic triangulation of $\sigma(\mu)$.

We will assume that a choice of surfaces $\sigma(\mu)$ has been
made, and examine the dynamical behaviour of the global variables
$M(\mu), B(\mu)$.

The Poisson bracket algebra can be deduced from the bracket
$[B^A_{ij} (x)$,\break $A_k{}^B (x^\prime)] = \epsilon_{ijk} g^{AB}
\delta(x - x^\prime)$, using a lattice regularization, following the
same lines as in (2+1)-dimensional gravity [6]. One finds

$$[M^A_B(\mu) , \ M^C{}_D(\nu)] = 0\eqno(2.20)$$

$$[B^A (\mu), \ M^C{}_D (\nu)] = \delta_{\nu\mu} C^{AC}{}{}_E
M^E{}_D(\mu) \eqno(2.21)$$

$$[B^A(\mu), \ B^B(\nu)] = \delta_{\mu\nu} C^{AB}{}{}_C \
B^C(\mu),\eqno(2.22)$$

\noindent
where $C^A{}_{BC}$ are the structure constants of $L(G)$. The
first bracket is deduced directly from $[A_i{}^A (x),
A_j{}^B (x^\prime)] = 0$. The second can be derived from the
expressions (2.14) and (2.16), by considering the contribution
from the point $x = x^\prime = P$, where $\gamma(\mu)$ intersects
$\sigma(\mu)$. The bracket of $B^A(\mu)$ with $B^B(\mu)$ is
ill-defined because the same surface $\sigma(\mu)$ appears on
both sides of the bracket; equation (2.22) is the result of the
lattice regularization mentioned above.

There are two sets of constraints on the global variables
$\{M(\mu), B(\mu)\}$ . The former are the cycle conditions for
the representations $M:\pi_1 (\sigma(\mu)) \to G$, presented by a
subset of the matrices $\{M(\nu)\}$: for each surface $\sigma(\mu)$,
there is a contractible loop (the ``cycle''), which follows every
basis loop of $\pi_1 (\sigma(\mu))$ in both directions. If
$\sigma(\mu)$ is a genus $g$ surface and $\pi_1 (\sigma(\mu))$ is
presented by the basis $\{\gamma(\mu), \mu = \mu_1, \mu_2, \cdots
, \mu_{2g} \}$, the cycle conditions are

$$\eqalign{ W(\mu) = &\biggl( M(\mu_1) M^{-1}(\mu_2) M^{-1}(\mu_1)
M(\mu_2)\biggr) \biggl( M(\mu_3) M^{-1} (\mu_4) M^{-1} (\mu_3)\cr
& M(\mu_4)\biggr) \biggl(M(\mu_{2g-1}) M^{-1} (\mu_{2g-1})
M(\mu_{2g})\biggr) = I,\cr}\eqno(2.23)$$

\noindent
since the connection is flat and the cycle is homotopically
trivial. The conditions (2.23) are the remmants of the flatness
conditions for the global variables:

$$F_{ij}{}^A \approx 0 \longrightarrow W(\mu) - I \approx 0
\eqno(2.24)$$

\noindent
The other constraints are a consequence of the closure conditions
$D\wedge B = 0$, and of our definition of $B^A(\mu)$, based on
removing the region bound by $2\times dim H_2(\Sigma^3)$ basis
surfaces $\sigma_{\pm\epsilon}
(\mu)$, to obtain a topologically trivial open manifold
$\Sigma^3_\epsilon$. The integral of the two-form over the boundary,
in the limit $\epsilon \to 0$, is a sum over $\mu$ of the
expressions (2.16) and (2.17), and vanishes as a consequence of
the definitions (2.16), (2.17) and the closure of
$B^A{}_{ij}(x)$. Thus,

$$D\wedge B = 0 \longrightarrow J^A\equiv \Sigma_\mu (1 -
M^{-1}(\mu))^A{}_B B^B(\mu) \approx 0. \eqno(2.25)$$

The brackets (2.20) - (2.22), together with the constraints
(2.24) and (2.25), define a dynamical system in the constrained
Hamiltonian formulation, which we constructed to have the same
physical content as the $B\wedge F$ field theory. Since some
steps in this construction are non-trivial, particularly the
regularization of the brackets $[B^A(\mu), B^B(\mu)]$, we will
state this equivalence of physical content as a conjecture.

We will say that theories $A$ and $B$ are ``equivalent'' if there is
a gauge transformation which maps solutions of $A$ to solutions
of $B$, and this map is bijective.

\

\noindent
\underbar{Conjecture 2.1} The constrained Hamiltonian system given
by the brackets (2.20) - (2.22) and the constraints (2.24),
(2.25), with $H \approx 0$, is equivalent to that which derives
from the $B \wedge F$ field theory (2.1), for any
coordinate-independent choice of representation of the basis of
$H_2(\Sigma^3)$ in $Z_2 (\Sigma^3)$, $\sigma : \mu
\longrightarrow \sigma (\mu)$.

\

\noindent
2.3 Solution of the time-evolution problem.

\

The constraints (2.24) - (2.25) have a first-class algebra with
the following structure:

$$[J^A, J^B] = C^{AB}{}{}_C J^C \eqno(2.26)$$

$$[J^A, (W(\mu)-I)^B{}_C] = C^{AB}{}{}_D
(W(\mu)-I)^D{}_C - (W(\mu)-I)^B{}_D
C^{AD}{}{}_C\eqno(2.27)$$

$$[(W(\mu)-I)^A{}_B, (W(\nu)-I)^C{}_D] =
0. \eqno(2.28)$$

\noindent
The closure conditions $J^A \approx 0$ generate global
$G$-transformations:

$$[J^A, B^B (\mu)] = C^{AB}{}{}_C B^C(\mu)\eqno(2.29)$$
$$[J^A, M^B{}_C(\mu)] = C^{AB}{}{}_D M^D{}_C(\mu) - M^B{}_D(\mu)
C^{AD}{}{}_C \eqno(2.30)$$

\noindent
The cycle conditions $W^A_B (\mu) \approx I$ generate
transformations of $B^A(\mu)$, which include timelike
translations of the base-point $P$, at which all loops
$\gamma(\mu)$ and surfaces $\sigma(\mu)$ intersect. The time
evolution is generated by the Hamiltonian constraint

$$H = \sum_\mu Tr \biggl( \xi(\mu) (W(\mu) - I)\biggr),
\eqno(2.31)$$

\noindent
where $\xi(\mu)$ are generalized ``lapse-shift'' functions.
Note that, given (2.20) and (2.21),

$$\ddot B^A(\mu) = \left[ H, [H, B^A(\mu)]\right] =
0 ,\eqno(2.32)$$

$$\dot M^A{}_B (\mu) = [H, M^A{}_B (\mu) ] = 0 .\eqno(2.33)$$

\noindent
Thus, the integrated two-forms are linear functions of time
and the holono-\break mies are constants of the motion.

The number of physical degrees of freedom is equal to the number
of configuration space variables minus the number of first-class
constraints. There are $dim L(G)$ independent closure conditions
(2.25), but the cycle conditions are not all independent: they
are associated with surfaces $\sigma_\epsilon(\mu)$ which,
together with their partners $\sigma_{-\epsilon}(\mu)$, form the
boundary of a homotopically trivial manifold $\Sigma^3_\epsilon$; this
implies that the product of the cycle conditions over all faces
$\sigma_\epsilon (\mu)$ and $\sigma_{-\epsilon}(\mu)$ is an
identity, so that $dim L(G)$ of the cycle conditions are
redundant, and the number of independent cycle conditions is
$dim L(G) \times  (dim
H_2(\Sigma^3) - 1)$. Altogether, in terms of the Betti numbers
$b_i = dim H_i(\Sigma)$, one has $dim L(G) \times (b_1)$
configuration variables $M(\mu)$, minus $dim L(G) \times (1 + b_2
- 1)$ constraints, or

$$dim L(G) \times (b_1 - b_2) = 0 \eqno(2.34)$$

\noindent
degrees of freedom, using Poincar\'e's identity $b_i = b_{3-i}$.
Note that one could also count directly the degrees of freedom in
the representations $M: \pi_1(\Sigma^3) \to G_1$ as $b_1 \times dim
L(G)$ minus the number of independent cycle conditions, $(b_2-1)
\times dim L(G)$, minus $dim L(G)$ for the overall $G$-conjugacy.
Again, one finds zero degrees of freedom.

The ``generic'' counting which lead to (2.34) must be modified
for some topologies, for which there are fewer independent cycle
conditions. Non-contractible spheres in the set $\{\sigma(\mu)\}$
do not contribute any cycle conditions, and non-contractible tori
provide only the $dim (L(G)) - rank (L(G))$ independent conditions

$$[M(\mu_1), \ M(\mu_2)] = 0 \eqno(2.35)$$

If one considers the orientable compact topologies which admit
locally homogeneous structures; only the quotients of $H^3$ and
$S^3$ have no non-contractible tori or spheres, as far
as we know. The fact that the $B\wedge F$ theory has no degrees
of freedom in these cases is related to Mostow's rigidity theorem
on the discrete representations of $\pi_1 (\Sigma^3)$ into
semisimple Lie groups with trivial centers and no compact
factors, not isomorphic to $SL(2, R)$ [7].

We conclude this section by counting the physical degrees of
freedom when $\Sigma^3$ has the topology of any one of Thurston's
eight classes of locally homogeneous orientable compact manifolds
[8]. The results are summarized in Table 3.1 in the next section.

\

\noindent
1.\underbar{Type a: quotients of $E^3$}.

\

The topologies are $T^3, T^3/Z_2, T^3/Z_3, T^3/Z_4, T^3/Z_6$ and
$T^3/Z_2 \times Z_2$. The fundamental group $\pi_1(\Sigma^3)$ has
three generators which are dual to three non-contractible tori.
The cycle conditions require that the matrices $\{ M(\mu), \mu =
1, 2, 3\}$ commute. The closure conditions are then $r =$
rank$(G)$ times redundant, since the matrices $(1 - M(\mu))$ in
the adjoint representation have $r$ common null eigenvectors. If
we denote by $d = dim L(G)$ the dimension of the Lie algebra,
there are $3\times d$ variables, minus $2(d-r)$ cycle conditions
and $d-r$ independent closure conditions, or $3\times r$ physical
degrees of freedom ($6r$ phase space degrees of freedom).

\

\noindent
2. \underbar{Type b: quotients of Nil}.

\

$\Sigma^3$ is a non-trivial $S^1$-bundle over $T^2$. Two of
the cycles are as in case (1) above, but the third fixes
$M(3)$ as a function of the commutator of $M(1)$ and $M(2)$. For
example, for type $b_{LR} / 1 (n)$,

$$M(1) M^{-1}(2) M^{-1}(1) M(2) = M(3)^n. \eqno(2.36)$$

\noindent
This relation and one of the commutators can be chosen as a set
of $d+(d-r) = 2d-r$ independent cycle conditions. One easily
shows that they imply $M(3)^n = I$. If one considers the
``generic'' solution, $M(3)=I$, then there are only
$d-r$ independent closure conditions. Note that the holonomy
generates a non-faithful representation $\pi_1(\Sigma^3) \to G$ in
this case. The number of degrees of freedom is $3d - (2d-r + d-r)
= 2r$.

\

\noindent
3. \underbar{Type c: quotients of $H_2 \times R$}.

\

$\Sigma^3$ is finitely covered by a trivial $S^1 -$ bundle over
$\Sigma_g$, a genus $g$ Riemann surface. The cycle conditions are

$$[M(\mu), M(2g+1)] = 0 \ \ (\mu = 1, \cdots , 2g) ,\eqno(2.37)$$

$$\eqalign{ &(M(1) M^{-1}(2) M^{-1}(1) M(2)) (M(3) M^{-1} (4)
M^{-1}(3) M(4))\cr
& \cdots M(2g-1) M^{-1}(2g) M^{-1}(2g-1) = I \cr}\eqno(2.38)$$

If one chooses $M(2g+1) = I$, all equations (2.37) are satisfied
and, with (2.38), one has a total of $2d$ independent conditions.
Otherwise $(M(2g+1) \not= I)$, all of the $M(\mu)$ commute among
themselves and equation (2.39) is satisfied in a trivial way. In
either case, the representation $M: \pi_1 (\Sigma^3) \to G$ is
notfaithful. For $M(2g+1) = I$, one has $3d$ constraints and
$(2g-2)\times d$ degrees of freedom, while in the other case the number
of constraints is $(d-r)\times 2g + (d-r)$ and there are $r\times(2g+1)$
degrees of freedom.

\

\

\

\noindent
4. \underbar{Type d: quotients of $SL(2,R)$}.

\

$\Sigma^3$ is a non-trivial $S^1-$bundle over $\Sigma_g$. The
cycle conditions are (2.37) and one relation which fixes
$M(2g+1)$ as a function of the $\Sigma_g-$ cycle (2.38). In
the case $M(2g+1) = I$, one has $(2g-2)\times d$ physical
degrees of freedom. The representations $M: \pi_1(\Sigma^3) \to
G$ are not faithful.

\

\noindent
5. \underbar{Type e: quotients of $H^3$}.

\

There is an infinite set of compact quotients of $H^3$, such as the
polyhedra discovered by L\"obell [9] with two hexagonal faces and
twelve pentagonal faces identified in pairs; the classification
of these quotients is not complete. None of them are
smoothly deformable, as a consequence of Mostow's theorem [7].
This theorem also tells us that a basis of $H_2(H^3/\Gamma)$ can have
at most one non-contractible torus, since there
are no free parameters in the quotienting group $\Gamma: \pi_1(\Sigma^3)
\to G$, in accordance with the counting (2.34) (one of the
cycles is redundant, and could be of the torus type, but as far as
we know quotients of $H^3$ do not admit any
non-contractible tori or spheres in $Z_2 (H^3/\Gamma))$. Thus,
Mostow's theorem implies that the $B\wedge F$
theory has no degrees of freedom for the hyperbolic manifolds.

\

\noindent
6. \underbar{Type f: quotients of Sol}.

\

$\Sigma^3$ is finitely covered by a (non-trivial) $T^2-$ bundle
over $S^1$. Since the cycle is invariant under mapping class
transformations (Dehn twists), the cycle conditions are the same
as for type a topologies; so is the counting of degrees of
freedom of the $B\wedge F$ theory.

\

\noindent
7. \underbar{Type g: quotients of $S^3$}.

\

$\Sigma^3$ is one of the following [10]: $S^3, S^3/Z_m, S^3/D_m$,
where $D_m$ is the symmetry group of a regular m-gon in the plane, and
$S^3/T, S^3/O$ and $S^3/I$, where $T,O$ and $I$ are the symmetry
groups of the regular tetrahedron, octahedron and icosahedron in
$R^3$. The second holonomy group is trivial, and so is $\pi_1
(\Sigma^3)$: the $B \wedge F$ theory has no physical degrees of
freedom.

\

\noindent
8. \underbar{Type h: quotients of $S^2 \times R$}.

\

$\Sigma^3$ is finitely covered by $S^2\times S^1$. There are no
cycle conditions, but $b_1 = b_2 = 1$: the phase space is spanned
by a single matrix $M \epsilon G$ and an ``integrated two-form''
$B^A$ with values in $L(G)$. The closure conditions are

$$J^A = (I - M^{-1})^A{}_B B^B, \eqno(2.39)$$

\noindent
or $d-r$ independent conditions, leaving $r$ physical degrees of
freedom.

This completes the list of solutions of the $B\wedge F$ theory;
in the next section we examine the possible relation between
these solutions and flat spacetimes $\Sigma^3 \times (0,1)$, when
$G = SO(3,1)$.

\

\noindent
3. FLAT SPACETIMES.

\

If $B = e \wedge e$ and $G = SO(3,1),$ then $F = R$ is the Riemann
curvature and the action 2.1 becomes the Palatini action

$$S = \int e \wedge e \wedge R.\eqno(3.1)$$

This begs the question: When do the solutions of the $B \wedge F$
theory represent flat spacetimes? Clearly, any flat spacetime
gives a solution of the $B\wedge F$ theory, by $B = e \wedge e$.
The converse is not true, as we will see shortly. We will first
review the global properties of flat spacetimes, then see under
what conditions the global variables $\{ B(\mu), M(\mu)\}$ of the
$B \wedge F$-theory describe flat spacetimes.

\

\noindent
3.1 $ISO(3,1)$ holonomy and flat spacetimes.

\

Given a flat spacetime $M = \Sigma \times (0,1)$, where $\Sigma$
is a compact, orientable, spacelike three-manifold, the holonomy
injects $\pi_1(\Sigma)$ into $ISO(3,1)$. Following Mess [11], one
can classify the flat spacetimes by the rank of the kernel of
the linear holonomy. If the rank is equal to three (three
independent translations), then $\Sigma$ is a three-torus and the
maximal development is all of $R^{3+1}$ (Thurston type a
universe; see conjecture 3.1 below). If the rank is 2, then
$\Sigma$ is finitely covered by $T^3$ and the maximal development
is again $R^{3+1}$ (Thurston type b, f). If the rank is 1,
$\Sigma$ is a closed Seifert bundle with rational Euler class
zero. The universal cover of the maximal development is the
direct product of $R$ and a domain $W$ in $R^{2+1}$, which is the
universal cover of the maximal development of a (2+1)-dimensional
flat spacetime $\Sigma^{(2)} \times (0,1)$, where $\Sigma^{(2)}$
is a genus $g$ Riemann surface (Thurston type c and d universes).
If the rank is zero, then the holonomy embeds $\pi_1(\Sigma)$ into
$SO(3,1)$ as a discrete compact subgroup, which is rigid by
Mostow's theorem. These spacetimes are quotients of the interior
of the future light-cone by this subgroup, have an
orbifold-singularity at the origin and admit a natural foliation
by the hyperbolic surfaces $H^3 \equiv t^2 - x^2-y^2-z^2 =
\tau^2$ (Thurston type $e$ universes).

We will denote the holonomy for $\gamma(\mu)$ by $\{M(\mu),
b(\mu)\}$ where $M(\mu)$ is the $SO(3,1)$ projection of the
holonomy in the usual $4 \times 4$ matrix representation, and
$b(\mu)$ is the four-vector

$$b^a(\mu) = \oint_{\gamma(\mu)} M^a{}_b (s) e^b{}_i (s)
ds^i\eqno(3.2)$$

\noindent
where

$$M(s) = P exp \biggl\{-i \int_{\gamma(\mu)}^{(s)} A_i dl^i\biggr\} ,
\eqno(3.3)$$

\noindent
A flat spacetime $\Sigma^3 \times (0,1)$ can be represented by an
open subset of $R^{3+1}$ with points identified, ${\cal P} \times (0,1)$,
where ${\cal P}$ is a polyhedron with $2 \times dim H_2(\Sigma^3)$ faces which
are identified in pairs by the isometries $\{ M(\mu), b(\mu),
\mu = 1, \cdots , dim(H_2(\Sigma^3)) \}$. The edges of the polyhedron
are given by the vectors $b(\mu)$ and their images under the
$SO(3,1)$ components of the identifications, $M(\nu) b(\mu)$. All
corners of the polyhedron are identified to a single point of
$\Sigma^3$, which is the base point $P$ of the homotopy group
$\pi_1(\Sigma^3)$. To reconstruct the representation of the spacetime
${\cal P} \times (0,1)$, from the polyhedron ${\cal P}_0$ given by
$\{ M(\mu), b(\mu)\}$, one constructs a one-parameter family of
polyhedra by displacing one corner of ${\cal P}_0$ along a
 timelike segment $S(\tau), \tau \epsilon(0,1)$, and the other
corners along the identified segments $(M(\mu) S(\tau)$, etc.).
One can show that different choices of the timelike segment
$S(\tau)$ lead to representations of the same spacetime $\Sigma^3
\times (0,1)$. Likewise, the geometry of the polyhedral faces can
be chosen arbitrarily: It is sufficient to give the relative
positions of the corners, $b(\mu)$, and the $SO(3,1)$ holonomies
$M(\mu)$, to specify completely the spacetime. Roughly speaking,
a polyhedron $\cal P$ can be obtained from $\Sigma^3$ by chosing a
base point $P \epsilon \Sigma^3$, cutting out $dim H_2 (\Sigma^3)$
basis surfaces and ``unfolding'' the resulting three-manifold
into a polyhedron. The point $P$, and the basis surfaces,
correspond to different choices of segments $S(\tau)$ and
polyhedral faces $\sigma(\mu)$.

We will list all flat spacetimes, by Thurston type, and in each
case count the number of parameters needed to describe them. We
are using Thurston's classification as a list of topologies which
are likely to have some relevance in cosmology, regardless of
whether or not there exists a foliation of these spacetimes in
locally homogeneous leaves. If such a foliation exists for any
flat spacetime with topology $\Sigma^3 \times (0,1)$, where
$\Sigma^3$ is a compact, orientable three-manifold, then the
spacetimes given below are locally homogeneous cosmologies, and
the list is complete: Any flat spacetime $\Sigma^3 \times (0,1)$
belongs to one of the categories listed below. We conjecture that
a locally homogeneous foliation always exists:

\noindent
\underbar{Conjecture 3.1} Let $M$ be a flat spacetime with
topology $\Sigma^3 \times (0,1)$, where $\Sigma^3$ is a compact,
orientable three-manifold with orientable normal bundle. There
exists a foliation of $M$ such that the metric on each leaf,
which is inherited from the flat metric on $M$, is locally
homogeneous.

\

\noindent
1. \underbar{Type $a_1$ cosmologies}

\

The universe is a three-torus. The holonomies $\{ M(\mu), b(\mu);
\mu = 1,2,3 \}$ must satisfy the $SO(3,1)$ cycle conditions,
which state that the matrices $M(\mu)$ commute, and the following
polyhedron closure relations [Figure 3.1]

$$b(1) + M^{-1}(1) b(2) - M^{-1}(2) b(1) - b(2) = 0 \eqno(3.4)$$

$$b(1) + M^{-1}(1) b(3) - M^{-1}(3) b(1) - b(3) = 0 \eqno(3.5)$$

$$b(2) + M^{-1}(2) b(3) - M^{-1}(3) b(2) - b(3) = 0 \eqno(3.6)$$

$$[M(1), M(2)] = [M(1), M(3)] = 0 \eqno(3.7)$$

The most general solution to the cycle conditions (3.7) is given,
in the appropriate frame, by

$$M(\mu) = exp \biggl( \alpha(\mu) J_{xt} + \beta(\mu) J_{yz}
\biggr) \eqno(3.8)$$

In the generic case, when all six coefficients $\{ \alpha(\mu),
\beta(\mu); \mu = 1,2,3 \}$ are non-zero and different from each
other, the only solutions to (3.4) - (3.6) are polyhedra which
collapse to a point singularity at a finite proper time in the
past or future: at the singularity $b (\mu) =0$, and at time
$\tau$ one has

$$b(\mu) = \left( M^{-1} (\mu) - I \right) N \tau,
\eqno(3.9)$$

\noindent
where $N$ is a timelike normal.

The largest family of solutions is found when $\beta(\mu) = 0$,
for all $\mu$. In this case there is again an initial or final
singularity (unless $\alpha(\mu) = 0$ as well), but one where the
singular polyhedron collapses to the $y - z$ plane, which is
invariant under $M(\mu)$. The polyhedron at time $\tau$ is then
given by

$$b(\mu) = b_0(\mu) + \left( M^{-1}(\mu) - I\right) N \tau,
\eqno(3.10)$$

\noindent
where $b_0^x (\mu) = b^t_0 (\mu) = 0, \forall\mu.$ One needs
five scalar parameters to specify $\{ b_0(\mu)\}$ up to a rotation
in the plane, on top of the three independent parameters
$\alpha(\mu)$, altogether eight parameters to specify the
spacetime.

It is interesting to compare these spacetimes to the Kasner
solutions [12]

$$ds^2 = -dt^2 + \sum_i t^{2b_i} dx^2_i \eqno(3.11)$$

\noindent
which are flat if one chooses $b = (1,0,0)$. They are then
specified by six parameters, which define the identifications in
$R\hskip-10pt I \ ^3, \Gamma: \vec x \sim \vec x + n_1 \vec a +
n_2 \vec b + n_3 \vec c$ and the quotient $T^3 = R\hskip-10pt I \
^3/\Gamma$. At $\tau = 0$, the spatial part of the metric (3.11)
collapses to a parallelogram in the $y-z$ plane, comparable to
(3.10) but with $b_0(3) = 0$. The vanishing of these two
parameters in the Kasner solution is due to the requirement that
the $T^3$ universes be surfaces of instantaneity: this
requirement is not satified in the solution (3.10) when $b_0(3) =
0$.

\

\noindent
2. \underbar{Type $a_n, n > 1$}

\

For $\Sigma^3 = T^3/Z_2$, the two images of the surface
$\sigma(3)$ are identified after a twist by 180$^\circ$. This is
consistent with the closure of the polyhedron's faces only if
$b(1)$ and $b(2)$ are orthogonal to $b(3)$. With these two new
conditions, the largest set of spacetimes with Type $a_2$ slices
is parametrized by 6 numbers. The polyhedron is given, as for
Kasner's solution, by equation (3.10) with $b_0(3) = 0$.

For $T^3/Z_3$ and $T^3/Z_6$, the vectors $b(\mu)$ must have the
same lengths and equal bihedral angles (4 conditions). The flat
spacetimes with Type $a_3$ and $a_5$ sections are labeled by 4
independent parameters. Finally, the flat spacetimes with $a_4$
and $a_6$ sections are labeled by 2 parameters and three
parameters, respectively.

\

\noindent
3. \underbar{Type $b$ cosmologies}

\

$\Sigma^3$ is a non-trivial $S^1$-bundle over $T^2$, and has
negative sectional curvature. There are seven subclasses, all of
which can be represented as a hexahedron with identifications. We
will consider only the type $b_{LR}/1(n)$. The
cycle conditions are

$$[M(1), M(3)] = 0, \eqno(3.12)$$

$$[M(2), M(3)] = 0, \eqno(3.13)$$

$$M(1) M^{-1}(2) M^{-1}(1) M(2) = M(3)^n.\eqno(3.14)$$

\noindent
There are three constraints on the polyhedron vectors $b_0(\mu)$
in the solution (3.10), on top of $\alpha(3) = 0$ from (3.14)
(note, however, that the identification rules are not the same as
for $T^3$). This gives the largest set of spacetimes in this
class; they are parametrized by the variables $\alpha(1),
\alpha(2)$ and the vectors $b_0(\mu)$ which satisfy the
constraints, altogether four parameters.

\

\noindent
\underbar{Type $c$ cosmologies}

\

The universe is finitely covered by $\Sigma_g \times S^1$, where
$\Sigma_g$ is a genus $g$ surface. The cycle conditions were
given in (2.37) - (2.38):

$$[M(\mu), M(2g+1)] = 0, \eqno(3.15)$$

$$W(2g+1) = I, \eqno(3.16)$$

\noindent
and the face closure conditions are [Figure 3.2]:

$$\cases{ \biggl( I - & $M^{-1} (2g+1)\biggr)
b(1)$\cr
& $ + \biggl( M(1)M(2) M^{-1}(1) - I \biggr)
b(2g+1) = 0$ \quad \qquad \qquad \qquad \ (3.17) \cr
\biggl( I - & $M^{-1}(2g+1) \biggr) b(2)$ \cr
& $ + \biggl(
M(2)M^{-1}(1) - M(1)M(2)M^{-1}(1)\biggr)
b(2g+1) = 0$ \qquad \ (3.18)  \cr
 etc.\hfill & \cr}$$

If $M(2g+1) = I$ and $M(1), \cdots , M(2g)$ generate a
faithful representation $M: \pi_1 (\Sigma_g) \to SO(3,1)$, there
are $2g\times 4$ closure conditions and 12 independent conditions
on the matrices $M(\mu)$. If one subtracts the overall
equivalence by $ISO(3,1)$ conjugacy, one is left with $(2g+1)
\times 10 - 8g - 12 - 10 = 12g - 12$ parameters to describe these
spacetimes. This is also the number of parameters which describe
the flat $(2+1)$-dimensional spacetimes with the topology
$\Sigma_g \times (0,1)$. Indeed, we have set
$M(2g+1) = I$, so these spacetimes are just the
direct product of $(2+1)$-dimensional spacetimes with $S^1$.
There are other solutions to (3.15)-(3.18) besides the ones we
have just described, for $M(2g+1) \not= I$; as in
Sect. 2, these have fewer free parameters.

\

\noindent
\underbar{Type $d$ cosmologies}

\

The universe is finitely covered by a non-trivial $S^1$-bundle
over $\Sigma_g$; the cycle conditions and closure conditions are
as above, except for (3.16), which now sets $M(2g+1)$ as a
function of the $\Sigma_g$-cycle $W(2g+1)$ (equation 2.38):

$$W(2g+1) = M(2g+1)^n. \eqno(3.19)$$

\noindent
Since  $[M(\mu), M(2g+1)] = 0$ for $\mu =1, \cdots , 2g$, the
generic solution satisfies $W(2g+1) = M(2g+1) = I$,
which we have discussed in Type $e$ above, and again there is
a $(12g-12)$-parameter family of solutions. Note, however, that the
identifications of points on the boundary of the polyhedra are
different. Roughly speaking, by following the cycle on $\Sigma_g$
one goes around $S^1$ n times.

\

\noindent
\underbar{Type e: hyperbolic spaces}

\

The spacetime is a quotient of the interior light-cone by a
discrete subgroup of $SO(3,1)$. There is a natural foliation in
terms of the Lorentz-invariant hyperbolic surfaces $t^2 =
\tau^2 - x^2 - y^2 - z^2$. The discrete subgroups are not
completely classified to this date, and they have no free
parameters by Mostow's theorem. One may check this fact in our
formalism, for any given topology, by counting the constraints
and showing that
these are sufficient to fix all the variables $\{ M(\mu), b(\mu)
\},$ up to an overall $ISO(3,1)$ transformation.

\

\noindent
\underbar{Type f cosmologies}

\

The universe can be represented as a hexahedron, as $T^3$, but
where one face is identified to the opposite face my means of a
Dehn twist. The cycle conditions are unchanged (from $T^3$),
since the cycle is mapping-class invariant, but one must impose
three constraints on the vectors $\{ b(1), b(2), b(3)\}$, which
require that the faces which are identified with a Dehn twist,
say $\mu = 1$, can be cut diagonally into two similar isoceles
triangles, and lie in a plane orthogonal to the vector $b(1)$:

$$b(1) \cdot b(2) = 0 \eqno(3.20)$$

$$b(1) \cdot b(3) = 0 \eqno(3.21)$$

$$b^2(2) + 2(b(1) \cdot b(2)) = 0 \eqno(3.22)$$

The largest set of spacetimes in this class is parametrized by
8 - 3 = 5 numbers.

\

\noindent
\underbar{Type g: Spherical spaces}

\

There are no flat spacetimes with a foliation in spherical
locally homogeneous slices, $S^3/\Gamma$. Indeed, if there were such a
spacetime, one could map $\Sigma^3$ as a sphere embedded in
Minkowski space, with points identified. This is impossible
because $S^3$ cannot be embedded in $R \hskip-10pt I \ ^{3+1}$.

\

\noindent
\underbar{Type h: Kantowski-Sachs spaces}

\

There are no flat spacetimes with foliations in $S^2 \times
S^1$ leaves, i.e. the topology of the Kantowski-Sachs
solutions. One proves this as for the quotients of the
three-sphere (Type $g$), by noting that it is impossible to embed
$S^2 \times (0,1)$ in $R \hskip-10pt I \ ^{3+1}$ in such a way
that $S^2$ is spacelike and the segment (0,1) is timelike.

The results of this section are summarized in Table 3.1 and
compared to the counting of solutions of the $B \wedge F$ theory
and to the Teichm\"uller parameters of locally homogeneous
structures given in [13]. The number of spacetime parameters is
generally less than twice the number of Teichm\"uller
parameters, because arbitrary initial conditions in the
cotangent bundle to Teichm\"uller space generally do not lead to
flat spacetimes.

\

\noindent
3.2 Global Torsion.

\

Given a set of global variables $\{ M(\mu), B(\mu) \}$, we wish
to interpret the matrices $M(\mu)$ as the $SO(3,1)$ holonomies
for loops $\gamma(\mu)$ in a flat spacetime, and each $B(\mu)$ as
the area bivector of the face $\sigma(\mu)$ of a polyhedron $\cal
P$ in $R \hskip-10pt I \ ^{3+1}$, which represents a section of
the spacetime as explained in Sect. 3.1. First of all, note that
the area bivectors of a polyhedron in $R\hskip-10pt I \ ^{3+1}$
depend only on the boundary segments of the faces, and not on
their local metric properties. We denote the boundary segments of
an n-gon $\sigma(\mu)$ by $b_i(\mu), i = 1, \cdots , n,$ with
$\Sigma_i b_i(\mu) = 0.$
The area bivectors can be computed from these vectors by
triangularizing the n-gon and summing the contributions of each
triangle. For example, for a quadrilatteral face, $B(\mu) = 1/2
(b_1(\mu) \wedge b_2(\mu) + b_3(\mu) \wedge b_4 (\mu))$. One
would like to invert these expressions, to compute the polygon
vectors which lead to given bivectors $B(\mu)$. It is not always
possible to find such vectors in such a way that the polygon
faces close: we will define the global torsion for a
``polyhedron'' given by $b^a_i (\mu)$, by

$$T^a(\mu) = \sum_{i=1}^n b_i^a (\mu)\eqno(3.23)$$

We will show with a simple example how one can construct
solutions of the $B\wedge F$ theory which  cannot be represented
as torsion-free polyhedra. Consider the spacetime $T^3 \times (0,1)$,
with trivial $SO(3,1)$ holonomies: a static three-torus
cosmology. If one modifies this solution by taking $M(1)$ to be a
rotation in the plane of the face $\sigma(1)$, then the bivectors
$B(1)$ and $M^{-1}(1) B(1)$ do not change but the two identified
faces are twisted with respect to each other. On easily shows
that the other faces no longer close, regardless of the choice of
$\{ b(\mu)\}$ consistent with the bivectors $\{ B(\mu) \}$. Yet,
all of the constraints of the reduced $B \wedge F$ theory are
satisfied.

One can identify a set of constraints on the variables $\{
M(\mu), B(\mu) \}$ which guarantee that the global torsion
vanishes. We will give these constraints below with a brief
explanation; the reader is referred to analogous work in the
exact lattice formulation of the $B \wedge F$ theory for a more
detailed derivation [14].

The torsion constraints state that the pairs of faces which
intersect at an edge of the polyhedron intersect transversally,
and the various identified edges respect the identification rules
of the polyhedron. For the intersection of faces $(\mu)$ and
$(\nu)$, the former conditions (transversality) are

$$\epsilon_{abcd} \ B^{ab} (\mu) \ B^{cd} (\nu) = 0 \eqno(3.24)$$

\noindent
The latter are found by considering all sets of three faces which
intersect along two identified edges: Let $B(\mu)$ and $B(\nu)$
represent two faces which intersect along edge No. 1, and let
$B(\nu)$ and $B(\rho)$ intersect along edge No. 2, which is
identified to edge No. 1 by means of an $ISO(3,1)$ transformation
with the $SO(3,1)$ component $M$. To reconstruct the triple
intersection at edge No. 1, one transports $B(\rho)$ from edge No. 2
to edge No. 1, and considers the three bivectors $B(\mu), B(\nu)$
and $MB(\rho)$, which intersect transversally in pairs by (3.5).
They intersect along the same edge if

$$C_{ABC} B^A(\mu) B^B(\nu) M^C{}_D B^D (\rho) = 0, \eqno(3.25)$$

\noindent
where $C_{ABC}$ are the structure constants of $SO(3,1)$ and $M$
is in the adjoint representation.

\

\
\

\noindent
4. CONCLUSION

\

We have investigated two theories, the $B \wedge F$ gauge theory
and the reduction of Einstein's equations to flat spacetimes, and
examined the relation between them.

The $B\wedge F$ theory was reduced to a finite constrained
Hamiltonian system, where the phase space is spanned by
holonomies $M : \pi_1 (\Sigma^3) \to G$ and ``integrated
two-forms'', $B : H_2 (\Sigma^3) \to L(G)$. This reduced theory
could then be solved explicitly when $\Sigma^3$ is one of the
orientable topologies listed by Thurston in his classification of
geometric structures, and the number of degrees of freedom was
given in all cases.

The flat spacetimes were constructed from their $ISO(3,1)$
holonomies $\{ M(\mu), b(\mu) \}$ for a basis set of loops of
$\pi_1 (\Sigma^3)$, as a subset ${\cal P} \times (0,1)$ of $R\hskip
-10pt I \ ^{3+1}$ with points identified on the boundary faces of
the polyhedron $\cal P$. The vectors $b(\mu)$ give the
displacements between corners of $\cal P$, while the matrices
$M(\mu) \epsilon SO(3,1)$ are the $SO(3,1)$ holonomies for loops
which cross through identified faces of the polyhedron. The
number of parameters which label the flat spacetimes $\Sigma^3
\times (0,1)$ were given when $\Sigma^3$ is one of Thurston's
manifolds. This completes Ellis's list of topologically
non-trivial cosmological models in the case of vanishing
spacetime curvature [15]. The solutions of the $B \wedge F$
theory with $G = SO(3,1)$ can be interpreted as representing
spacetimes with global torsion, which we define as the failure of
the polyhedra's faces to close. The variables $B^{ab} (\mu)$
represent the area bivectors of the faces of a polyhedron, when
they do close. We also gave the vanishing torsion constraints as
equations on the bivectors $B^{ab} (\mu)$.

The solution of the time evolution problem for the reduced
$B\wedge F$ theory can be exploited to give information on the
global dynamics of flat spacetimes, when the torsion constraints
are satisfied. In particular, one has the result that the area
bivectors of non-contractible surfaces, in an intrinsically
chosen set of representatives of basis elements of $H_2
(\Sigma^3)$ in $Z_2 (\Sigma^3)$, have a linear evolution in time.

The derivation of the reduced theory from the $B\wedge F$ field
theory required some non-trivial steps, and we could only
conjecture that the reduction is exact. It would be of great
interest to prove this conjecture, perhaps as in
the polygon representation of $(2+1)-$dimensional gravity [16]
or, by means of the exact lattice
theory [6],[14]. Also, the polyhedron representation of flat
spacetimes is based on several assumptions, in particular it
assumes the existence of non-contractible geodesic loops which
become the polyhedron's edges $b(\mu)$ in the Minkowski space
representation. If this construction can be formalized, it would
give a generalization of Poincar\'e's polygon in $H^2$ [16], which
labels the conformal structures on Riemann surfaces, to polyhedra
in $R \hskip-10pt I \ ^{3+1}$ which label the flat spacetimes.

Finally, the existence of a finite constrained Hamiltonian
formulation of the $B \wedge F$ theory could be useful in
developping the corresponding quantum theory [17]. The linear

time-evolution of the
classical variable $B(\mu)$ indicates that the quantum theory may
have a representation in terms of ``free particles'' in $L(G)$, but
with identifications given by the mapping class transformations,
or ``large diffeomorphisms'' [18]. The
Green's function could then be constructed by the method of
images, as a sum of freely propagating amplitudes over the
mapping-class images of the source. The quantum effects would
likely appear as interference terms in this sum, as in
(2+1)-dimensional quantum gravity [19].

\vfill\eject

\noindent
\underbar{Acknowledgements:}

\

I would like to express my gratitude to the organizers of the
ITP workshop on the small scale structure of spacetime, where
many of the results were derived thanks to numerous discussions
with Akio Hosoya, Steve Carlip, Vince Moncrief, Gary Horowitz,
as well as e-mail exchanges with Geoffrey Mess. The global
torsion constraints are the result of work done in collaboration
with Jos\'e Antonio Zapata.

\vfill\eject
\noindent
\underbar{References:}

\

\item{[1]} G.T. Horowitz, ``Exactly soluble diffeomorphism
invariant theories'', Commun. Math. Phys. 125 (1989), 417-437.

\item{[2]} E. Achucarro and P. Townsend, ``A Chern-Simons action
for three-dimensional anti-de Sitter supergravity theories, Phys.
Lett. B180 (1986), 89.

\item{[3]} E. Witten, ``2+1 gravity as an exactly soluble
system'', Nucl. Phys. B311 (1988), 46.

\item{[4]} P.A.M. Dirac, ``Lectures on Quantum Mechanics''
(Graduate School of Science Monographs Series) (New York: Belfer
1964).

\item{[5]} C. Teitelboim, Phys. Lett. B167 (1986), 63.

\item{[6]} H. Waelbroeck, ``2+1 Lattice Gravity'', Class. Quantum
Gravity 7 (1990), 751;

\item{ \ } ``Time-dependent solutions of 2+1 gravity'', Nucl.
Phys. B364 (1991), 475-494.

\item{[7]} G.D. Mostow, ``Strong Rigidity of Locally Symmetric
Spaces'', Ann. Math. Studies 78. (Princeton Univ. Press 1973).

\item{[8]} W.P. Thurston, ``Three-dimensional Manifolds, Kleinian
Groups and Hyperbolic Geometry'', Bull. Amer. Math. Soc. 6
(1982), 357-381.

\item{ \ } P. Scott, ``The Geometries of Three-manifolds'', Bull.
London Math. Soc. 15 (1983), 401-487.

\item{[9]} F. Lobell, Ber. Verhand. S\"acks. Akad. Wiss. Leipsig,
Math. Phys., Kl., 83 (1931), 167.

\item{[10]} J.A. Wolf, ``Spaces of Constant Curvature''. (Mc.
Graw-Hill, New York 1967).

\item{[11]} G. Mess, Lorentz Spacetimes of Constant Curvature'':
Institut des \break Hautes Etudes Scientifiques preprint IHES/M/90/28
(1990).

\item{[12]} C. Misner, K. Thorne and J.A. Wheeler:
``Gravitation'' (W.H. Freeman, San Francisco, 1970).

\item{[13]} T. Koike, M. Tanimoto and A. Hosoya: ``Compact
Homogeneous Universes'', Tokyo Institute of Technology Preprint
TIT/HEP-208/COS MO-26 (1992).

\item{[14]} H. Waelbroeck and J.A. Zapata, ``Topological Lattice
Gravity'', Mexico Preprint ICN-UNAM-93-13.

\item{[15]} G.F.R. Ellis, ``Topology and Cosmology'', Gen. Rel.
Grav. 2 No. 1 (1971), 7-21.

\item{[16]} H. Poincar\'e, ``Th\'eorie des Groupes Fuchsiens'',
Acta. Math. 1 (1982), 1-62

\item{ \ } B. Maskit, ``On Poincar\'e's Theorem for Fundamental
Polygons'', Adv. Math. 7 (1971), 219-230

\item{ \ } H. Waelbroeck, ``The Polygon Representation of Flat
(2+1)-dimensional Spacetimes'', Rev. Mex. Fis. 39 Vol. 6 (1993).

\item{[17]} E. Martinec, ``Soluble Systems in Quantum Gravity'',
Phys. Rev. 30D No. 6 (1984), 1198-1204.

\item{[18]} J.L. Friedman and D.M. Witt, ``Homotopy is not
Isotopy for Homeomorphisms of 3-Manifolds'', Topology 25 No. 1
(1986), 35-44

\item{ \ } J.L. Friedmann and D.M. Witt, ``Problems on
Diffeomorphisms Arising from Quantum Gravity'', Contemporary
Mathematics 71 (1988), 301-310.

\

\

\

\item{[19]} G. 't Hooft, Comm. Math. Phys. 117 (1988), 685.

\item{ \ } S. Carlip, ``Exact Quantum Scattering in 2+1
gravity'', Nucl. Phys. B324 (1989), 106-122.

\item{ \ } S. Carlip, ``The Modular Group, Operator Ordering, and
time in (2+1)-dimensional Gravity'', University of California
Preprint UCD-92-23 (gr-qc/9209011)

\item{ \ } H. Waelbroeck, ``Quantum Gravity in 2+1 Dimensions'',
Mexico Pre-\break print ICN-UNAM-93-14.

\vfill\eject

\noindent
Figure Captions

\vskip1cm

\noindent
\underbar{Fig. 3.1} A $T^3$-universe is represented as a
hexahedron in $R \hskip-10pt I \ ^{3+1}$ with opposite faces
identified.  As one transports a vector from the face $\sigma(\mu)$
to its identified partner $\sigma(- \mu)$, the vector is Lorentz-
transformed by the
matrix $M^{-1}(\mu)$. The closure of each face is a condition on
the basis vectors $b(\mu)$ and the matrices $M(\mu)$; for instance
the closure of the forward face requires that $b(1) + M^{-1}(1)
b(2) - M^{-1}(2) b(1) - b(2) = 0.$

\vskip2cm

\noindent
\underbar{Fig. 3.2} A universe with topology $\Sigma_2 \times
S^1$, where $\Sigma_2$ is a genus $g = 2$ surface, is
represented as a decahedron in $R \hskip-10pt I \ ^{3+1}$ with
the ``top'' and ``bottom'' octagons identified by $M(2g+1)$, and
the $4g$ ``latteral'' faces identified in pairs. The vector
$b(2g+1)$ at $A$ is identified to $M^{-1}(1) b(2g+1)$ at $D$, to
$M(2) M^{-1}(1) b(2g+1)$ at $C$ and finally $M(1) M(2) M^{-1}(1)
b(2g+1)$ at $B$, so the face $B(1)$ closes if
$(I - M^{-1}(2g+1) b(1) + (M(1)M(2)M^{-1}(1) - I)
b(2g+1) = 0.$

\vskip2cm

\noindent
\underbar{Table. 3.1} The number of parameters which label the
solutions of the $B \wedge F$ theory and the flat spacetimes
$\Sigma^3 \times (0,1)$ are given, when $\Sigma^3$ is one
of the Thurston geometries.  The rank of the algebra L(G) is $r$
and its dimension is $d$; for $SO(3,1)$, $r = 2$ and $d = 6$.

\end